\documentstyle[twocolumn,prb,aps,epsf]{revtex}

\begin{document}
\input epsf

\wideabs{
\title{Vortex-loops and solid nucleation in superfluid $^4$He and $^3$He}
\author{Nir Gov}
\address{Department of Physics,\\
University of Illinois at Urbana-Champaign,1110 Green St., Urbana 61801,
U.S.A.}

\maketitle
\tightenlines
\widetext
\advance\leftskip by 57pt
\advance\rightskip by 57pt

\begin{abstract}
We propose a new model for the nature of the nucleation of solid from the superfluid phases of $^4$He and $^3$He.
Unique to the superfluid phases the solid nucleation involves an extremely fast solidification front.
This results in a local release of pressure and a velocity field in the superfluid.
The superfluid velocity in turn facilitates the nucleation of vortex-loops.
The kinetic energy gain of this process balances the surface tension, as the solid surface is quickly covered by many vortex-loops ("hairy snow-ball").
We show that this scenario gives good agreement with experiments on heterogeneous nucleation, which differ with the classical theory of homogeneous nucleation by 8 orders of magnitude.
We propose several experiments that could show the involvement of vortices with solid nucleation.
\end{abstract}

\vskip 0.2cm
}

\narrowtext
\tightenlines
\section{Introduction}

The problem of nucleating a solid out of the superfluid phase of $^4$He is long-standing. The challenge is to explain the large reduction in the energy barrier from the homogeneous nucleation theory (HNT) value of $\sim 10^8$K, to the measured value of $10-50$K. The HNT energy barrier is huge due to the large surface tension energy $\sigma_{sl}\sim 0.2$erg/cm$^2$ and small density difference between the solid and liquid phases ($\sim 10$\%).
The HNT describes a static energy balance between the bulk and surface energies of a spherical buble of one phase inside the other
\begin{equation}
F(R)=4\pi R^{2} \sigma_{sl} -\frac{4}{3}\pi R^{3} \Delta G
\label{hnt}
\end{equation}
where $R$ is the radius of the solid and $\Delta G=\delta P (\rho_{s}-\rho_{l})/ \rho_{l}$.
From this the critical radius and energy barrier are calculated, for any given over-pressure $\delta P$. For $\delta P\sim$mbar we have $R_{c}(HNT)\sim1\mu$m and $\Delta E(HNT)\sim10^8$K.
Various suggestions have been put forward to attribute the large discrepancy with experiment as being due to surface impurities \cite{balibar00}.

A solid nucleating out of the fluid at an over-pressure $\delta P$ (above the melting pressure), will cause a local release of this pressure, due to the larger density of the solid phase.
This release of pressure causes a flow field in the fluid. 
In a normal fluid the finite viscosity and slow heat diffusion lead to slow dendritic growth.
By contrast, in the superfluid the pressure release can result in large fluid velocities and fast solid growth.
The dynamic effects of the flow field created in the superfluid by a solid nucleation where not treated previously. We propose here a model where this fluid flow allows the reduction of the free energy barrier of the nucleation process.

The release of the overpressure $\delta P\sim3-30$mbar at the solidification front results in a fluid velocity of: $v=\sqrt{2\delta P/\rho}\sim 2-6$m/sec.
These calculated velocities are supported by the following experimental observations.
High solidification velocities are found in experiments of solid growth (after nucleation), where a dislocation-free crystal grows in bursts of very high velocity \cite{ruutu1,tsymbalenko}. 
The liquid velocity is related to the solidification front velocity $v_{s}$ by: $v=v_{s}(\rho_{s}-\rho_{l})/ \rho_{l}\simeq 0.1 v_{s}$, where $\rho_{s}$ and $\rho_{l}$ are the solid and liquid densities.
When the growth is due to moving surface steps, these are found \cite{ruutu1} to have a velocity of $v_{s}\sim300$m/sec, i.e. close to the velocity of sound.
The velocity of these steps therefore coresponds to local fluid velocity of $\sim10$m/sec.
The nucleation events themselves \cite{ruutu,tsepelin} are characterized by extremely fast ($<0.5$msec) release of the overpressure, again indicating fast dynamics at the nucleation site.
The results of these experiments indicate the very likely occurence of high fluid velocity at the solidification front, in superfluid $^4$He and $^3$He.

We will show that these velocities are high enough to nucleate vortex-rings (Fig.1). The process of vortex-loop nucleation and growth due to a velocity field was previously studied in critical velocity experiments through micro-apertures \cite{avnel}.
We propose that the solid nucleation is helped by the simultaneous nucleation of vortex-loops on the solid surface.

\section{Thermally activated vortex-solid nucleation in superfluid $^4$He}

The expression for the energy of a half-vortex-loop at a solid boundary is given by
\begin{equation}
E_{vl}=\frac{1}{4}\rho \kappa_{4}^2 R \eta+E_{core}
\label{evl}
\end{equation}
where $R$ is the vortex-loop radius, $\rho$ is the superfluid density, $\kappa_{4}=h/m$ and $\eta=1/2 \ln (1+R^2/a^2)$ is the regularized logarithmic factor, with $a$ the effective vortex-core radius.
The core-energy $E_{core}\simeq \pi^2 R a_{0}^2 E_{cond}\rho/m$, where $a_{0}$ is the bare core radius \cite{core} and $E_{cond}$ is the superfluid condensation energy per atom, which is of order $\sim1$K in superfluid $^4$He and $\sim1$mK in superfluid $^3$He.
In superfluid $^4$He the core energy $E_{core}$ (\ref{evl}) is small since $a_{0_4}\sim1.4$\AA, and is usually neglected \cite{avnel}, while in superfluid $^3$He we have $a_{0_3}\sim120$\AA \ so this energy is important (see section V).
 
On a boundary substrate the effective core radius $a$ is larger than $a_{0_4}$: $a\sim4-50$\AA, as regards the dynamics of the vortices \cite{avnel,williams}.
This is due in part to small cavities on the surface, where there is a high density of vortices \cite{pinning}, and a lower barrier for vortex-loop nucleation. We describe these static effects by an increase of the effective core radius $a$ in (\ref{evl}). 
We will therefore follow the critical velocity experiments \cite{avnel}, and describe the surface influence on vortex-loop nucleation through this {\em single free parameter} $a$.
We note here that a hydrodynamic description (\ref{evl}) at length scales of a few \AA \ is suspect, and our ignorance of the exact behavior is hidden in the free parameter $a$.
Dynamic effects that change $a$ are discussed in section VII.

The vortex loop energy is reduced in the presence of a velocity field $v$
\begin{equation}
E_{vl}(v)=E_{vl}-{\bf P_{vl}}\cdot {\bf v}
\label{evlv}
\end{equation}
where the vortex-loop impulse is given by
\begin{equation}
P_{vl}=\frac{1}{2}\pi \rho \kappa_{4} R^2
\label{pvl}
\end{equation}
In Fig.1 we show possible orientations of the vortex-loop around the solid surface so that the local fluid velocity field $v$ lowers the vortex-loop energy. 

Using the relation $v=\sqrt{2\delta P/\rho}$, we plot in Fig.2 the energy barrier from Eq.(\ref{evlv}) as a function of overpressure $\delta P$. We assume ${\bf P_{vl}}\cdot {\bf v}=Pv$ in (\ref{evlv}), i.e. the maximum reduction in $E_{vl}$ due to $v$.
We find that for $a\sim10$\AA \ the critical overpressure where the barrier $E_{vl}=0$ (\ref{evlv}): $\delta P_{c}\sim30$ mbar. 
Above this critical overpressure the local fluid velocity at the nucleating solid surface makes the energy of a vortex-loop negative.
We also plot the different terms in (\ref{etot}) as a function of the overpressure.

We now have to consider how does the creation of vortices compete with energy cost due to the solid-liquid surface tension.
The overall energy balance of the nucleating solid is therefore
\begin{equation}
E_{tot}(R,\delta P)=N_{vl}(R)E_{vl}(R,\delta P)+E_{surf}
\label{etot}
\end{equation}
where $E_{surf}=4\pi R^2 \sigma_{sl}$ is the surface tension energy (\ref{hnt}).
We estimate that the number of half-vortex-loops that can cover the hemispherical solid nucleus is of the order: $N_{vl}\sim (R/a_{0_4})^2$ ("hairy snow-ball", Fig.1).
In Fig.2 we plot the total energy calculated simply by (\ref{etot}), as a function of overpressure, for $a=10$\AA.
We see that the critical overpressure at which $E_{tot}=0$ is $\delta P_{c-tot}\sim32$ mbar, slightly larger than the value for the single vortex-loop nucleation.
A factor of 10 reduction in the number of nucleated vortices on the solid surface, will increase $\delta P_{c-tot}$ by $\sim15$\% compared with $\delta P_{c}$ for a single vortex-loop.

As the number of vortices is very large ($\sim10^2-10^3$) the assumption that they do not interact is not true, and as a result their energy is not simply additive, as in (\ref{etot}).
Our inability to give a detailed description of the multi-vortex-loop flow on the solid surface does not prevent us from comparing with experiment, for the following reason.
We propose that it is the single vortex-loop energy barrier $E_{vl}$ (\ref{evlv}) which controls the thermal and quantum nucleation process. The reason is that once a single vortex-loop is "plucked" out of the wall, in conjunction with the solid nucleus, it will quickly multiply to cover the entire solid surface (Fig.3), due to the strong velocity field.
We presume that in the regime where the vortex-loop energy is negative, it is advantageous for a single nucleated vortex-loop to multiply and cover the nucleating solid.
The kinetic energy gain of this process balances the surface tension, and the solid is nucleated, covered by many vortex-loops ("hairy snow-ball").

A more detailed microscopic picture of the vortex-loop-solid interface is proposed in Fig.4.
The point where the vortex-loop touches the solid can be associated with a surface defect, such as a step.
The circular motion of this linear defect can be synchronized with the superfluid motion around the vortex 
so that the vortex feeds the moving solidification step. 
The kinetic energy of the fluid at the solidification surface, due to the release of the overpressure $\delta P$, is usually lost as the liquid becomes a stationary solid. Through the spontaneous vortex nucleation above the threshold (\ref{evlv}), this kinetic energy
drives the motion of the solidification step.

This dynamic scenario for the nucleation and fast growth of the solid, eventually slows down and a static solid of radius $R>R_{c}(HNT)$ remains.

\section{Comparison with experiments in $^4$He}

Experiments \cite{balibar,ruutu,sasaki} on solid nucleation from superfluid $^4$He have found that the nucleation rate increases with increasing over-pressure, saturating with a critical over-pressure parameter $\delta P_{c}$ (\ref{power}) of $\sim5$ mbar \cite{ruutu} and $\sim20$ mbar \cite{sasaki}.
It was found that above some temperature $T_{q}\sim0.2-1$K (see section IV) the nucleation rate is temperature dependent
\begin{equation}
w(T)=\Gamma \exp(-E_{b}/k_{B}T)
\label{nucrate}
\end{equation}
where $\Gamma$ is the attempt rate, which was found \cite{ruutu} to be $\Gamma\simeq k_{B}T_{q}/\hbar\sim10^{10}$Hz, where $T_{q}$ is the quantum nucleation temperature (see section IV).
The barrier energy for nucleation $\Delta E$ was found to have a $3/2$ power-law behavior 
\begin{equation}
\Delta E\simeq E_{0}(1-\delta P/\delta P_{c})^{3/2}
\label{power}
\end{equation}
with \cite{ruutu} $E_{0}\simeq50$K, and \cite{sasaki} $E_{0}\simeq30$K.

The main difference between the two experiments is that in the experiments by Ruutu {\em et. al.} \cite{ruutu}
the nucleation is on the copper substrate of the cell wall.
The nucleation was found to take place in unique and stable sites, which changed randomly upon total melting of the solid.
The nucleation in the experiments by Sasaki at. al. \cite{sasaki} takes place on an electrode, placed in the bulk of the superfluid, where an electric field creates a pulse of overpressure.

In Fig.5 we plot the critical overpressure $\delta P_{c}$ as a function of the core-radius parameter $a$.
We also plot the number of vortex-loops at the critical overpressure, as a function of the core radius parameter $a$. We find that a typical number is of order $\sim 10^2$.
Finally, we plot the radius $R_{c}$ of the solid nucleus at the critical overpressure (Fig.5).
We also note that
close to the critical overpressure, the calculated energy barrier for a single vortex-loop ($E_{vl}$) can be approximated as \cite{sasaki,ruutu} in Eq.(\ref{power}) (Fig.2).
This allows us to calculate the value of $E_{0}$ and compare to the experimental results (Table 1).

We note that the nucleation in the Ruutu et. al. experiment \cite{ruutu} is determined by prefered sites on the copper substrate, where the barrier for vortex nucleation is supposedly lowest.
Such pinning of vortices by solid substrates is well known \cite{pinning}.
We can take this static effect into account by an effectively larger core-radius parameter $a$ (Fig.4), compared to the Sasaki et. al. experiment.
If we compare our calculations for effective core radii $a\simeq 10$\AA \ and $25$\AA, to the experimental results of Sasaki et. al. \cite{sasaki} and Ruutu et. al. \cite{ruutu} respectively, we find 
that while in both cases we matched the measured  $\delta P_{c}$, the calculated $E_{0}$ is approximately a factor of 2-3 larger than measured (see Table.1).
This level of agreement is reasonable, considering our ignorance of the details of the flow and vortex tangle at the solidification front.
In any case it is a huge improvement over the HNT which differs with the experiment by $\sim 8$ orders of magnitude.

The similarity between the solid-nucleation phenomena and vortex-nucleation \cite{avnel} extends also to the statistical spread of the nucleation events.
Let us compare the spread of the critical overpressure with the spread in critical velocity at a micro-orifice \cite{avnel}.
In the critical velocity experiments of Avenel et. al. \cite{avnel} it was found that the vortex-loop nucleation had a spread in the critical velocity, of $\Delta v_{c}/v_{c}\simeq 0.04$ at $T=450$ mK.
It follows from our model that we should find the same spread in the solid nucleation experiments, at the same temperature.
We therefore expect a half-width of the nucleation probability, around the critical overpressure, of $\sim0.1$ mbar, in very good agreement with the measured distribution \cite{ruutu} at $T=500$ mK. 

\section{Quantum vortex-solid nucleation in superfluid $^4$He}

In both sets of solidification experiments \cite{ruutu,sasaki} it was found that below some temperature $T_q$ the nucleation rate becomes temperature independent, which may signal a regime of quantum nucleation by tunneling.
In the experiments by Sasaki et. al.\cite{sasaki} this temperature may be estimated as $T_{q}>\sim1$K, and in the experiments by Ruutu et. al.\cite{ruutu} this temperature was found to be $T_{q}\sim0.2$K.

The quantum nucleation temperature of $T_{q}\sim0.2$K found by Ruutu et. al.\cite{ruutu} is very close to that found by Avenel et. al. \cite{avnel} in critical-velocity experiments. In both these experiments the vortex-loops were found to nucleate at preferred sites on the boundary substrate.

Crudely, the temperature $T_{q}$ can be taken as the attempt frequency for vortex-loop nucleation $\Gamma$ in (\ref{nucrate}), and is found to be of the order of the "cyclotron" frequency of the vortex-loops (see discussion in Ref.\cite{core},ch.8.5):
$\Omega_{cyc}=\hbar/ma^2$. 
This amounts to the statement that zero-point quantum fluctuations of the vortex are dominant over the thermal fluctuations below $T_{q}=\hbar\Omega_{cyc}/k_{B}$.
The same expression is derived by calculating the zero-point energy of the helical waves in a vortex-loop of length $R_{c}$: $E_{zpe}\sim\hbar^2/mR_{c}^2$. 
Over the entire surface of the solid nucleus there are $N_{vl}$ sites for vortex-loop nucleation, so the overall attempt rate is: $E_{zpe}N_{vl}\sim\hbar\Omega_{cyc}$.
The classical hydrodynamic treatment of the vortices in section II amounts to integrating out the quantum fluctuations of the vortex core \cite{gordon}.

The calculated quantum temperatures $T_{q}$ for the two effective core radii of 10\AA \ and 25\AA \ are 0.12K and 0.02K respectively. These values are both a factor of $\sim$10 smaller than the quantum temperatures found in the experiments cited above.
This discrepancy of factor 10 is of the order of extra (logarithmic) factors that are usually found in vortex-waves experiments \cite{hall}.
The ratio of the calculated quantum temperatures agrees with the experimental measurements, leading us to conclude that the attempt frequency of the quantum and thermal nucleation $\Gamma$ is driven by zero-point oscillations of the vortex-loops. 
This attempt frequency then determines the thermal to quantum "transition" temperature \cite{core} $T_{q}$.

\section{Vortex-solid nucleation of superfluid $^3$He}

The experimental data \cite{normalhe3} concerning the solidification of $^3$He from the normal fluid indicates a classical behavior of slow growth, controled by diffusion and viscosity.
On the other hand, the solidification of superfluid $^3$He, shows activated nucleation, similar to the case of superfluid $^4$He,
with a critical overpressure of \cite{tsepelin} $\sim10-50$ mbar. 

The main difference between the superfluid $^3$He and $^4$He is that in superfluid $^3$He the vortex core-radius is \cite{corehe3}: $a_{0_3}\simeq120$\AA.
This means that the core energy $E_{core}$ is now an important factor (\ref{evl}).
We find that for $E_{cond}\sim3$mK and $a=a_{0_3}$ in (\ref{evl}) the critical overpressure is $\delta P_{c}\simeq 10$ mbar, of order of the measured data.
We take the surface tension in $^3$He as \cite{dobbs}: $\sigma_{sl-3}\sim 0.06$erg/cm$^2$. 
Fitting the barrier energy to a power-law (\ref{power}) near the critical overpressure, we find: $E_{0}\simeq600$K. 
The critical radius is found to be at its lower limit of the bare core-radius $a_{0_3}$, so that the "hairy snow-ball" picture is qualitatively different, with one vortex-loop engulfing the solid nucleus.

The fluid velocity we calculate corresponding to the critical overpressure is $\sim5$m/sec, which is larger than the critical velocity for pair-breaking \cite{parts} $v_{cp}=\hbar/ma_{0_3}\sim1$m/sec.
This means that the fluid at the nucleating solid surface is entirely normal.
This is in accordance with our finding that the entire solid nucleus is engulfed by the core of a single vortex-loop.

\section{Anomalous fast growth of $^4$He crystals}

Recent experiments \cite{ruutu1,tsymbalenko} have shown that independent of the nucleation of the solid, there is a transition in superfluid $^4$He
to a fast (anomalous) growth mode.
This transition occurs above a temperature dependent critical overpressure, as seen in Fig.6.
In this section we propose to explain this transition along the same lines we have advanced so far,
i.e. the interplay between the velocity field at the solid surface and vortex growth in the superfluid.

The surface of the hcp phase has a roughening transition chcracterized by a
Kosterlitz-Thouless transition of step-defects \cite{fisher}.
We propose that these can become associated with superfluid vortex-loops, and allow fast solid growth (Fig.4).
The density of defects on the 2D surface below the KT transition, is given by \cite{achilles}:
$n=(n_{s}/2)|T/(T_{KT}-T)|$, where $n_{s}\simeq 1/d^2$, $d\sim500$\AA \ is the scale of the surface defect separation (a free parameter in this model).

The two dimensional defects (steps) of the KT theory can behave as spiral growth centers, similar to screw dislocations. This happens if a fluid velocity parallel to the solid sustains such a spiral step motion, i.e. when vortex-loops are associated with such defects, as shown in Fig.4. 
It is only under these conditions, of vortex-loop nucleation (and growth), that these defects become growth centers and the fast growth mode is established.

We now ask what is the critical over-pressure above which the fluid velocity due to the released overpressure is larger than that required for the nucleation and growth of vortex-loops with areal density $n$.
From the definition of the vortex-loop energy (\ref{evl},\ref{evlv}) and momentum (\ref{pvl}) 
we have the critical fluid velocity
\begin{equation}
v_{crit}=\frac{E_{vl}}{P_{vl}}\simeq \frac{1}{2}\frac{\pi\kappa_{4}\eta}{R}
\label{vcrit}
\end{equation}
where we neglected $E_{core}$ in $^4$He.
We take the vortex core radius $a=a_{0_4}$ since the vortices are not on a substrate so that pinning effects are absent. The vortex-loop radius $R$ is the distance between surface defects of the solid, i.e. $R=1/\sqrt{n}$.

The critical over-pressure corresponding to this critical velocity $v_{crit}$ is (following the discussion in sections I-II)
$\delta P_{sv}=(1/2)\rho v_{crit}^2$.
In Fig.6 we plot this critical overpressure, and we find that it corresponds very well with the line separating slow and burst-like solid growth in the experiments \cite{ruutu1,tsymbalenko}.
In this calculation we used $T_{KT}=0.9$K in the equation for $n$, 
which corresponds to the second roughening transition of the a-facets $T_{R2}$.

We shall now calculate the velocity of solid growth in the normal direction to the solid surface (Fig.4).
The fluid velocity around the vortex-loop parallel to the solid surface, at the radius $R$ is $v_{fl}\sim v_{crit}$. The solid step that rotates around with the vortex, therefore has the velocity 
$v_{s}=v_{fl}\rho_{l}/(\rho_{s}-\rho_{l})\simeq 10v_{fl}$.
The time it takes the step to complete one rotation around the vortex is $R/v_{s}$ so that the normal growth velocity is given by $v_{n}=v_{s}(a_{s}/R)$, where $a_{s}\sim3$\AA \ is the step height (Fig.4).
In Fig.7 we plot this growth velocity, which we find to be in the 1-2m/sec range for temperatures 0.4-0.8K, which means a crystal of millimeter size after a burst of $\sim$msec, in agreement with the observations \cite{tsymbalenko}.

At this stage we do not give a description of the observation of a time delay in the slow-to-fast growth transition \cite{tsymbalenko}.

\section{Non-equilibrium effects}

The rate of change of overpressure ${\dot {\delta P}}=\partial \delta P/\partial t$ is not zero in any real experiment. The bigger this parameter the further we are from the thermodynamic equilibrium situation, for which the HNT applies.
The importance of ${\dot {\delta P}}$ in solid nucleation is emphasized by recent experiments \cite{balibar01},
that show the solidification of $^4$He due to short converging acoustic pulses. The critical overpressure for soldification was found to be $\sim4.7$bar at T=65mK. This value of the critical overpressure is much larger than those found before \cite{balibar,ruutu1,sasaki}.
The rate of growth of overpressure ${\dot {\delta P}}\simeq10^7$bar/sec in this last experiment is also much larger than in previous experiments \cite{ruutu1,sasaki}: $\sim10^{-4}$bar/sec and $\sim10^{-2}$bar/sec respectively \cite{ratesasaki}.
The trend of increasing nucleation barrier with increasing rate of change of the relevant thermodynamic parameter, i.e. overpressure in our case , is well known in other phase nucleation phenomena \cite{mixtures}.
Using the data from all three sets of experiments we find $\delta P_{c}\propto {\dot {\delta P}}^{0.26\pm 0.05}$. This is close to the exponent $\sim 1/3$ found in $^3$He-$^4$He phase seperation \cite{mixtures}, for example.

The occurence of the exponent $1/3$ can be qualitatively understood from the following.
In the time-dependent scaling theory of phase separation \cite{nigel}
it is found that in an intermediate asymptotic regime, the nucleated domain
has length $L\propto t^{1/3}$, where $t$ is the time since the system was moved away from the equilibrium line of the two phases.
The static critical size of the final stable domain is given by the HNT (\ref{hnt}) $R_{c}(HNT)=2\sigma/(\delta P (\delta \rho/\rho))$.
Equating this radius with the time-dependent domain size, we find that the critical overpressure for the stable static domain behaves as: $\delta P_{c}\propto t^{-1/3}\Rightarrow \delta P_{c}\propto {\dot {\delta P}}^{1/3}$.

The measured \cite{balibar01} critical overpressure of a few bars can be approximately matched in our model by taking an effective core size $a\sim a_{0_4}$ in (\ref{evl}) (which gives $\delta P_{c}\simeq1.5$bar).
Our model gives for such a case an energy barrier parameter of $E_{0}\simeq9$K (Table.1), in agreement with the measurements \cite{balibarqfs01}.
The data from all these experiments indicates that the rate of change of overpressure ${\dot {\delta P}}$ influences the dynamics of the nucleating vortices. We can describe this influence by changing the effective core radius $a$. The values in Table 1 show that ${\dot {\delta P}}$: $a\propto {\dot {\delta P}}^{-0.13}$, i.e. $\delta P_{c}\propto a^{-2}$. 
The last relation follows naturally from $\delta P_{c}\propto v_{fl}^2$ and $v_{fl}\propto 1/a$ (see Eq.\ref{vcrit}).

We note that our effective core radius parameter $a$ has to account for static (such as preferred pinning sites) and dynamic (${\dot {\delta P}}$) differences between the experiments. We therefore describe only a simplified model of the real dynamics of the superfluid and nucleating/growing solid.

\section{Conclusion}

Before concluding let us briefly compare with the process of $^3$He-$^4$He phase seperation at oversaturation, from the superfluid $^4$He phase \cite{mixtures}. The energy released upon phase separation is $\sim (\mu_{3}-\mu_{4})\Delta x$ per atom, where $\Delta x$ is the over saturation and $\mu_{3}-\mu_{4}\sim 0.1$K/atom.
This energy is available as kinetic energy of the superfluid $^4$He surrounding the nucleated $^3$He-rich buble. The resulting fluid velocity is $\sim1$m/sec for an oversaturation of $\Delta x \sim 1$\%. This fluid velocity corersponds to the case of solid nucleation at an overpressure of a few mbar, found in the solid nucleation experiments of Ruutu {\em et. al.} \cite{ruutu}, described above.
Moreover, the crossover from thermally activated to quantum (i.e. temperature independent) nucleation occurs at a similar temperature of $\sim100$mK. These features seem to suggest that large velocity gradients and therefore vortex-assisted nucleation may occur also at $^3$He-$^4$He phase separation.

We shall now propose several experiments that can check our model.
Critical velocity experiments through micro-orifices performed close to the melting pressure, have found that the phase-slip events are correlated with the appearance of a solid plug in the orifice \cite{erikqfs01}. This again is an indirect evidence for the connection between vortex and solid nucleation.
An indication for the occurence of vortices at the solid nucleation site, may come by the trapping of negative ions (injected electrons) in superfluid $^4$He. Below a temperature of $\sim1.7$K these negative ions are trapped at the cores of vortices \cite{core}, and these trapped charges might show that vorticity is created at the solid nucleation site.
In the case of solidification of superfluid $^3$He, the vorticity may be directly seen through NMR measurements \cite{he3nmr}.

To conclude, we have presented here a model where the dynamics of the superfluid in the form of quantized vortices is driven and in turn facilitates, the nucleation and growth of a solid nucleus.
This process may be unique to both superfluid $^3$He and $^4$He since the solid growth is very fast, with correspondingly fast and dissipation-less fluid velocities at the solidification front.
The model we presented is largely phenomenological, since we lack a microscopic theory that describes the dynamic solid nucleation event.
The proposed mechanism may be important in other nucleation phenomena in superfluids, such as the A-B transition in superfluid $^3$He.

{\bf Acknowledgement} 
I thank Tony Leggett and Sahng-Kyoon Yoo for many useful discussions and suggestions.
I also thank Nigel Goldenfeld for pointing out the time-dependent scaling relation.
This work was supported by
the Fulbright Foreign Scholarship grant,
NSF grant no. DMR-99-86199  and
NSF grant no. PHY-98-00978.

\begin{table}
\caption{The measured and calculated values of the critical overpressure $\delta P_{c}$ and energy barrier parameter $E_{0}$ (\ref{power}).}
\begin{tabular}{c|cc|c|cc}
Experiment&$\delta P_{c}$(exp.)&$E_{0}$(exp.)&$a$&$\delta P_{c}$(calc.)&$E_{0}$(calc.)\\ 
 &mbar&K&\AA&mbar&K \\ \hline
Ruutu et. al. \cite{ruutu}&5&50&25&5&160 \\ \hline
Sasaki et. al. \cite{sasaki}&20&30&10&31&60 \\ \hline
Balibar et. al. \cite{balibar01,balibarqfs01}&4000&10&1.4&1600&9 \\
\end{tabular}
\end{table}

\begin{figure}[tbp]
\centerline{\ \epsfysize 7cm \epsfbox{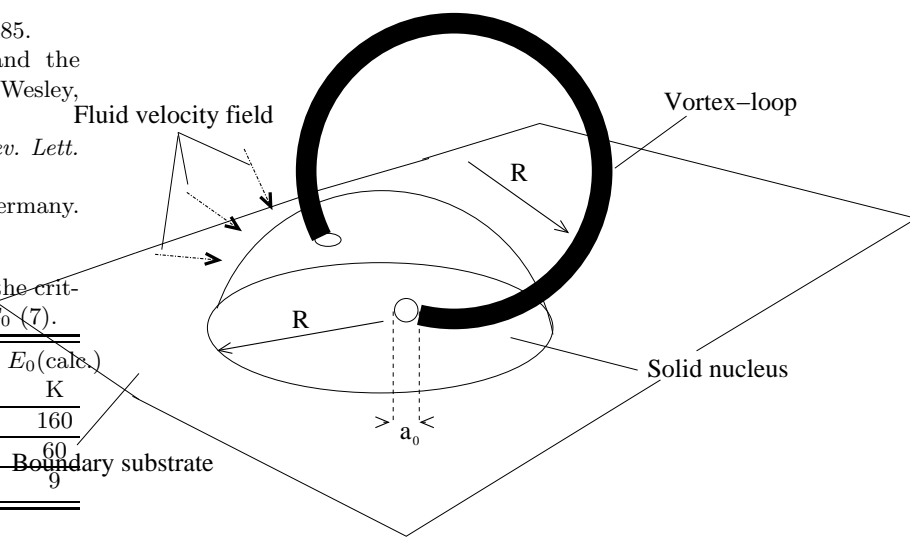}}
\vskip 3mm
\centerline{\ \epsfysize 10cm \epsfbox{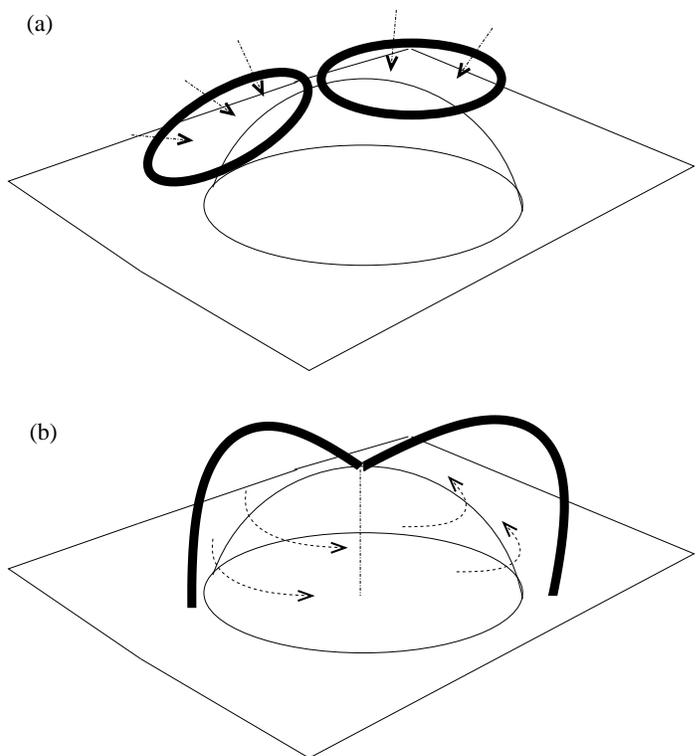}}
\caption{A schematic picture of the "hairy snow-ball" of vortex-loops on the growing solid nucleus.
(a) Possible arrangements of the vortex-loops, such that the local velocity field lowers the vortex-loop energy. In (b) we show a solid growing by a screw-dislocation (vertical dash-dot line), causing lateral fluid motion.}
\end{figure}

\begin{figure}[tbp]
\centerline{\ \epsfysize 7cm \epsfbox{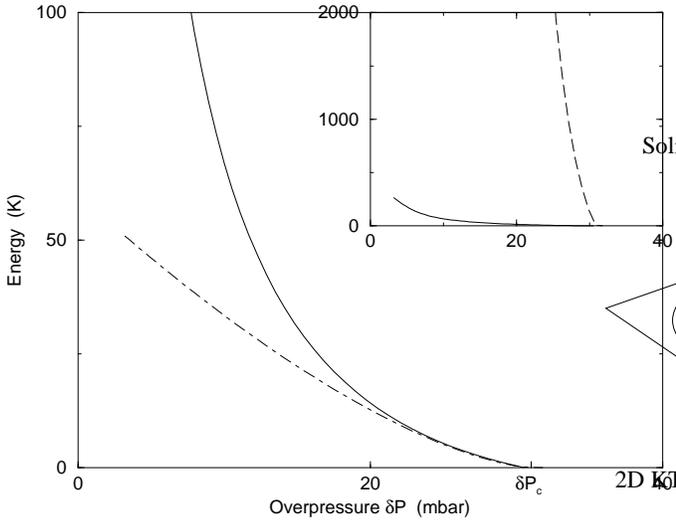}}
\caption{The vortex-loop barrier energy $\Delta E$ as a function of overpressure for effective core-radius $a=10$\AA. The inset shows the energy barrier of the solid nucleus including the surface tension (\ref{etot}) (dashed line). We also plot (dash-dot line) the polynomial fit to $\Delta E$, with $\delta P_{c}=31$ mbar and $E_{0}=60$K.} 
\end{figure}

\begin{figure}[tbp]
\centerline{\ \epsfysize 10cm \epsfbox{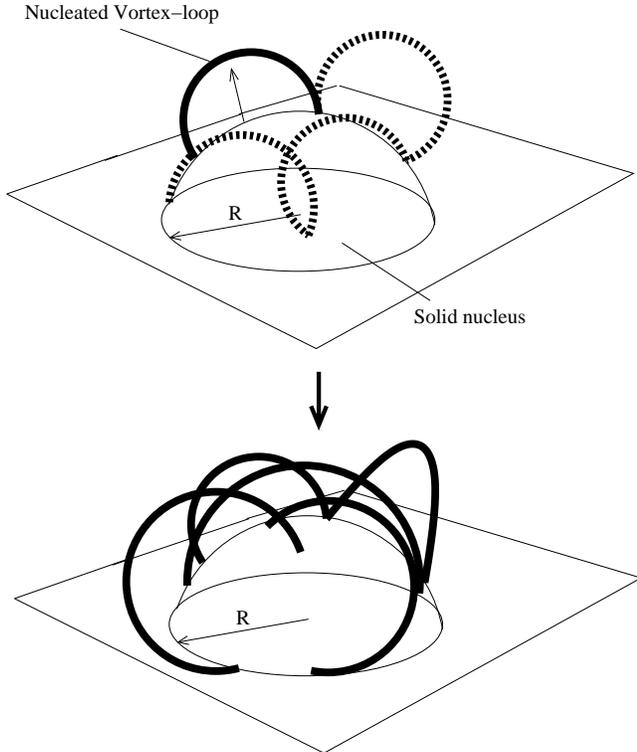}}
\caption{A schematic picture of how "plucking" a single vortex-loop over the energy-barrier given by (\ref{evlv}) $E_{vl}(v)$, leads to proliferation of the loops and creation of the "hairy snow-ball".} 
\end{figure}

\begin{figure}[tbp]
\centerline{\ \epsfysize 7cm \epsfbox{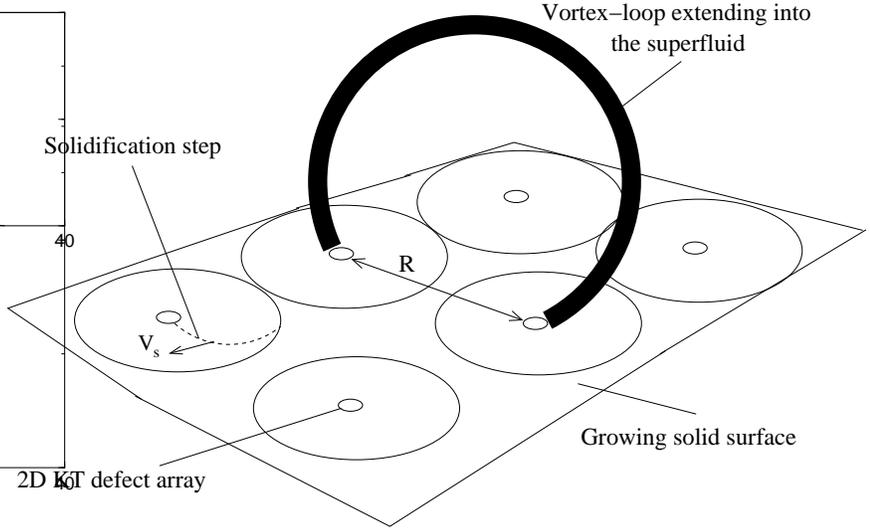}}
\caption{Superfluid vortices and solid defects (steps) at the superfluid-solid interface, leading to fast growth.}
\end{figure}

\begin{figure}[tbp]
\centerline{\ \epsfysize 7cm \epsfbox{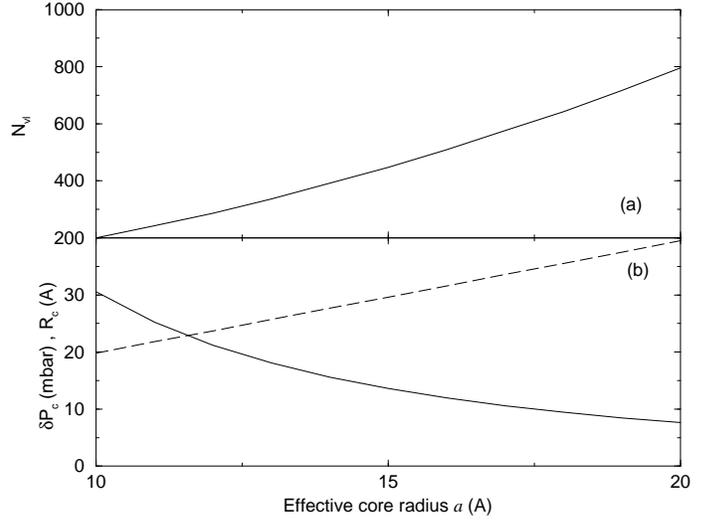}}
\caption{(a) The number of vortex-loops $N_{vl}$ at the (b) critical overpressure $\delta P_{c}$ (solid line), as a function of effective core-radius $a$. In (b) we also plot the critical radius $R_{c}$ (dash line).}
\end{figure}

\begin{figure}[tbp]
\centerline{\ \epsfysize 7cm \epsfbox{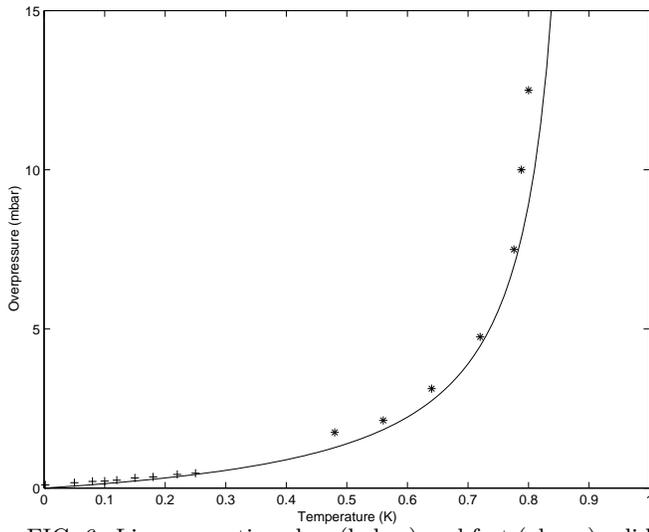}}
\caption{Line separating slow (below) and fast (above) solid growth in superfluid $^4$He. Experimental data: stars \cite{tsymbalenko} and crosses \cite{ruutu1}. Solid line- calculated critical overpressure $\delta P_{sv}$.}
\end{figure}

\begin{figure}[tbp]
\centerline{\ \epsfysize 7cm \epsfbox{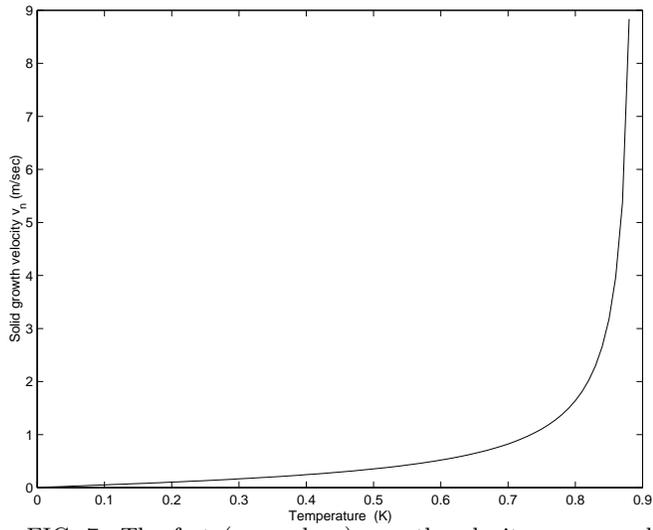}}
\caption{The fast (anomalous) growth velocity $v_{n}$ normal to the solid surface.}
\end{figure}


\begin{references}
\bibitem{ratesasaki} In the experiment of Sasaki et. al. \cite{sasaki} ${\dot {\delta P}}$ is not specified so we use the minimum pulse width of 10msec.
\end{references}
\end{document}